\documentclass[11pt]{article}

\usepackage[final]{acl}

\usepackage{times}
\usepackage{latexsym}

\usepackage[T1]{fontenc}

\usepackage[utf8]{inputenc}

\usepackage{microtype}

\usepackage{inconsolata}

\usepackage[most]{tcolorbox}

\usepackage[table]{xcolor}
\usepackage{tikz}
\usepackage{nicematrix}

\usepackage[utf8]{inputenc}
\usepackage{booktabs,multirow}
\usepackage{enumitem}
\usepackage{kotex}
\usepackage{color}
\usepackage{rotating}
\usepackage{amsmath}
\usepackage{amsfonts}
\usepackage{pifont}
\usepackage{array}
\usepackage{amssymb}
\usepackage{algpseudocode}
\usepackage{algorithm}
\usepackage{float}
\usepackage{multicol}
\usepackage{lipsum}
\usepackage[normalem]{ulem}
\usepackage{xcolor}
\usepackage{soul}
\usepackage{listings}
\usepackage{amsmath}
\usepackage{adjustbox}
\usepackage{graphicx} 
\usepackage{colortbl} 
\usepackage[dvipsnames]{xcolor}   
\usepackage{mdframed}
\usepackage{hyperref}
\usepackage{url}
\usepackage{xurl}

\tcbuselibrary{listings} 

\newcommand{\eg}{{\it e.g.}}

\title{From Relevance to Authority: Authority-aware Generative Retrieval in Web Search Engines}

\author{
    Sunkyung Lee$^1$\thanks{\ \ Equal contribution}, 
    Jihye Back$^2$\footnotemark[1], 
    Donghyeon Jeon$^2$, 
    Soonhwan Kwon$^2$, \\
    \textbf{Moonkwon Kim}$^2$\textbf{,} 
    \textbf{Inho Kang}$^2$\textbf{,} 
    \textbf{Jongwuk Lee}$^1$\thanks{\ \ Corresponding author} \\
    $^1$Sungkyunkwan University, Republic of Korea, 
    $^2$Naver Corporation, Republic of Korea \\  
    $^1$\texttt{\{sk1027, jongwuklee\}@skku.edu}, 
    $^2$\texttt{\{1oojihye, donghyeon.jeon, soonhwan2.kwon,} \\
    \texttt{moonkwon.kim, once.ihkang\}@navercorp.com}
}

\begin{document}
\maketitle
\begin{abstract}

Generative information retrieval (GenIR) formulates the retrieval process as a text-to-text generation task, leveraging the vast knowledge of large language models. However, existing works primarily optimize for relevance while often overlooking document trustworthiness. This is critical in high-stakes domains like healthcare and finance, where relying solely on semantic relevance risks retrieving unreliable information. To address this, we propose an \textit{\textbf{Auth}ority-aware \textbf{G}enerative \textbf{R}etriever (\textbf{AuthGR})}, the first framework that incorporates authority into GenIR. AuthGR consists of three key components: (i) \textit{Multimodal Authority Scoring}, which employs a vision-language model to quantify authority from textual and visual cues; (ii) a \textit{Three-stage Training Pipeline} to progressively instill authority awareness into the retriever; and (iii) a \textit{Hybrid Ensemble Pipeline} for robust deployment. Offline evaluations demonstrate that AuthGR successfully enhances both authority and accuracy, with our 3B model matching a 14B baseline. Crucially, large-scale online A/B tests and human evaluations conducted on the commercial web search platform confirm significant improvements in real-world user engagement and reliability.

\end{abstract}

\section{Introduction} 

\begin{figure}[t]
\centering
\includegraphics[width=1.0\linewidth]{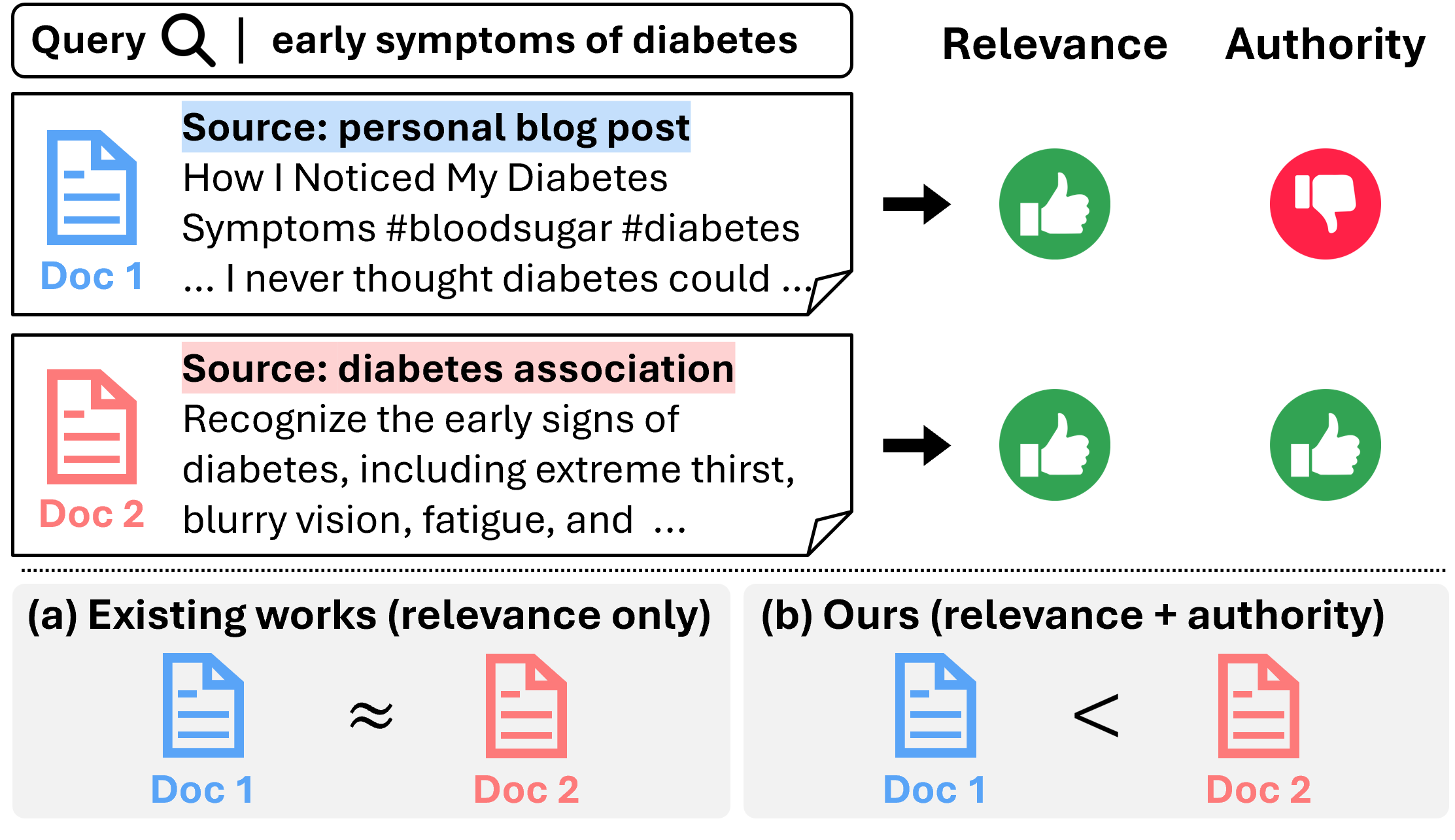}
\vspace{-3.5mm}
\caption{Illustration of our motivation. (a) Models relying solely on relevance fail to distinguish an unreliable blog from authoritative ones. (b) By integrating authority, our model can prioritize trustworthy documents.}\label{fig:motivation}
\vspace{-4.5mm}
\end{figure}

Generative Information Retrieval (GenIR) has emerged as a promising paradigm for the retrieval task, driven by recent advances in large language models (LLMs)~\cite{sigir/MetzlerTBN21/Rethinking}. Unlike traditional methods that encode queries and documents into vector representations~\cite{EMNLP/KarpukhinOMLWEC20/DPR}, GenIR reformulates retrieval as a text generation task~\cite{nips/Tay/DSI, sigir/Zeng24PAG}. It aims to directly generate \emph{document identifiers (DocIDs)} that satisfy users’ information needs. Recently, GenIR has also gained much attention in industrial applications beyond academic research, \eg, e-commerce search~\cite{www/Pang25GRAM, icdm/Shen24HiGen}, food delivery~\cite{acl/Zhang25HierGR}, and financial services~\cite{sigir/Shen25Alipay}, boosting both accuracy and user satisfaction.

Despite these advancements, existing GenIR methods primarily optimize semantic relevance, overlooking the critical dimension of the \textit{document authority}. This poses risks in high-stakes domains like healthcare and finance, where trustworthiness is essential. As illustrated in Figure~\ref{fig:motivation}, a relevance-only model may rank an unverified  personal health blog (Doc 1) as highly as an official medical association (Doc 2) solely due to topical similarity (Figure~\ref{fig:motivation}a). To prevent exposing users to potentially inaccurate or unverified information, it is essential to move beyond relevance and incorporate authority, ensuring the model to prioritize authoritative sources (Figure~\ref{fig:motivation}b).

However, explicitly integrating document authority into GenIR faces three challenges. (i) \emph{Defining authority}: Textual cues alone often fail to distinguish trustworthy sources from sophisticated promotional websites, making it difficult to define authority at scale. (ii) \emph{Learning authority}: Instilling the subtle and complex concept of authority without compromising semantic relevance is non-trivial, requiring training methods beyond standard fine-tuning. (iii) \emph{Deploying the authority-aware model}: Finally, simply replacing existing production rankers is impractical and risky. Instead, the model must be seamlessly integrated into current pipelines of a large-scale search platform, enhancing trustworthiness maintaining relevance.

To this end, we propose an \textit{\textbf{Auth}ority-aware \textbf{G}enerative \textbf{R}etriever (\textbf{AuthGR})}, which explicitly incorporates the document authority into the generative retrieval model. AuthGR consists of three key components: (i) We introduce \textit{Multimodal Authority Scoring} to establish a scalable and quantifiable definition of authority. This component employs a vision-language model to assess site trustworthiness using textual and visual signals, effectively automating human-like judgments. (ii) To effectively learn the abstract concept of authority, our model is trained via a progressive \textit{Three-stage Training Pipeline}. The process begins with domain-continued pre-training for foundational knowledge of the search domain, followed by supervised fine-tuning to learn the core task of DocID generation. The \textit{group relative policy optimization (GRPO)}~\cite{deepseek2024deepseek-math-grpo} stage lastly refines the model, explicitly training it to prefer high-authority documents. (iii) Finally, AuthGR is integrated using a \textit{Hybrid Ensemble Pipeline} with existing rankers, enabling seamless real-world deployment without compromising relevance.

To summarize, our contributions are as follows:
\begin{itemize}[leftmargin=*,topsep=0pt,itemsep=-1ex,partopsep=2ex,parsep=1ex]
\vspace{0.5mm}
\item \textbf{First authority-aware GenIR}: To our knowledge, this is the first work to systematically incorporate authority into GenIR, addressing the need for trustworthiness.

\vspace{0.5mm}
\item \textbf{Comprehensive training framework}: We introduce a three-stage training strategy that progressively instills the concept of authority using multimodal scores and preference optimization.

\vspace{0.5mm}
\item \textbf{Real-world impact}: AuthGR is thoroughly validated across three complementary evaluations: it matches a 14B baseline that is 4.7$\times$ larger in offline settings; it boosts user engagement in large-scale online A/B tests; and it achieves superior quality ratings in human evaluations.
\end{itemize}
\section{Related Work}

\begin{figure*}[t]
\centering
\includegraphics[width=1.0\linewidth]{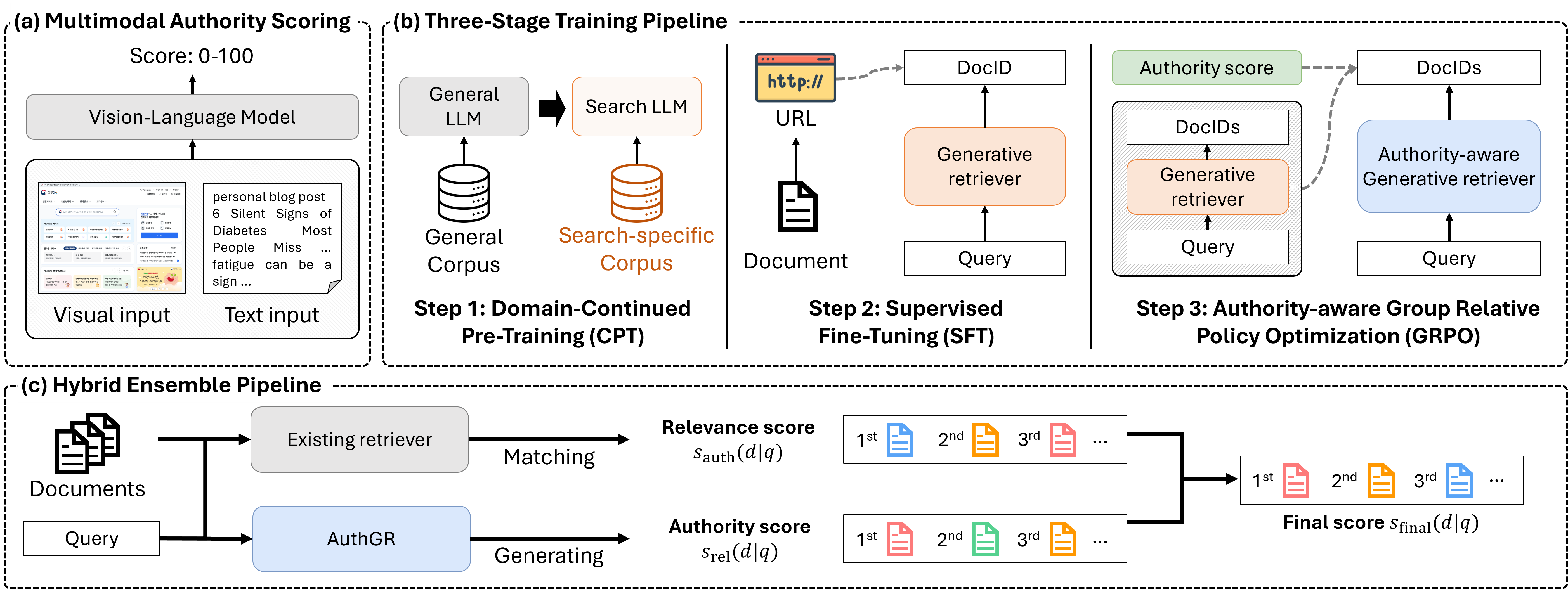}
\vspace{-4.5mm}
\caption{Overall architecture of AuthGR. (a) \textit{Multimodal Authority Scoring} quantifies document trustworthiness based on textual and visual signals. (b) \textit{Three-stage training pipeline} progressively instills authority awareness into the retriever. (c) The \textit{Hybrid Ensemble Pipeline} integrates AuthGR with existing rankers for deployment.}\label{fig:overall_architecture}
\vspace{-4.5mm}
\end{figure*}

\subsection{Generative Information Retrieval}
Generative Information Retrieval (GenIR) directly generates document identifiers (DocIDs)~\cite{sigir/MetzlerTBN21/Rethinking, SurveyGenSearchRecSys}. Early research explored diverse DocIDs, such as numeric IDs~\cite{nips/Tay/DSI}, URLs~\cite{acl/RenLWWW23/TOME}, or semantic keywords~\cite{emnlp/L23GLEN}. Subsequent work shifted towards optimizing ranking quality via  ranking-aware loss~\cite{aaai24LTRGR, sigir/Mekon25DDRO} or reinforcement learning~\cite{emnlp23GenRRL, acl25Synthetic}. Despite advancements, GenIR predominantly focuses on \textit{semantic relevance}, overlooking \textit{document authority}. This is critical as GenIR expands into industrial domains, \eg, finance~\cite{sigir/Shen25Alipay, sigir24AlipayLLMGR} or e-commerce~\cite{www/Pang25GRAM}, where prioritizing relevance over authority exposes users to potentially inaccurate content. See Appendix~\ref{app:related_work} for broader discussion.

\subsection{Trustworthiness in Information Retrieval}
Traditional information retrieval has long emphasized \textit{trustworthiness} by quantifying \textit{authority} via link structures such as PageRank~\cite{BrinP98PageRank} and HITS~\cite{Kleinberg99HITS}, as well as content signals like TrustRank~\cite{vldb/GyongyiGP04TrustRank, pvldb/Dong15KnowledgebasedTrust}. Recent initiatives such as the TREC Health Misinformation Track~\cite{trec/Clarke20MisInformation, trec/ClarkeMS21MisInformation} further underscore its necessity in high-stakes domains, requiring diverse credibility indicators including source expertise~\cite{sigir/ZhangTAS22TREC}. Despite this rich history, GenIR systems have yet to adapt these sophisticated assessment methods. Our work bridges this gap by systematically incorporating authority signals into generative retrieval, ensuring reliability for real-world deployment.

\section{Proposed Method}\label{sec:method}

We propose an \textit{\textbf{Auth}ority-aware \textbf{G}enerative \textbf{R}etriever (\textbf{AuthGR})}, the first framework to systematically integrate authority into GenIR. As depicted in Figure~\ref{fig:overall_architecture}, AuthGR comprises three key components: (i) \textbf{Multimodal Authority Scoring} to quantify trustworthiness via textual and visual cues; (ii) a \textbf{Three-Stage Training Pipeline} to progressively instill authority awareness; and (iii) a \textbf{Hybrid Ensemble Pipeline} for robust deployment.

\subsection{Multimodal Authority Scoring}\label{sec_authority_score} 
To quantify document authority at scale, we propose Multimodal Authority Scoring that automates human-like assessment. Traditional systems rely on fragmentary signals like link structures~\cite{BrinP98PageRank}, which often fail to capture trustworthiness holistically. In contrast, human judgment integrates textual content, visual design, and advertisement patterns. To replicate this intuition, we employ a Vision-Language Model (VLM) as a scalable proxy for human evaluators.

Concretely, the VLM jointly processes: (i) \textit{textual signals} including document title, body text, and URL metadata; and (ii) \textit{visual signals} from a page-level screenshot. This integration is vital since promotional content often mimics authoritative language, making text alone deceptive. Visual cues, \eg, ad intrusiveness and layout quality, are essential to distinguish genuine credibility from sophisticated mimicry, as illustrated in Appendix~\ref{app:vlm_example}. We prompt the VLM using a comprehensive rubric that evaluates the three core dimensions of Expertise, Officialness, and Public Interest, further supplemented by checks for commercial intent and harmfulness (\eg, spam, illegal content)\footnote{The full prompt is in Appendix~\ref{app:vlm_prompt}.}. Formally, the authority score for a document $d$ is defined as:
\begin{equation}\small
\texttt{Authority}(d) = f_{\text{VLM}}(T(d), V(d)) \in [0, 100],
\end{equation}
where $f_{\text{VLM}}$ is the scoring function from the VLM, and $T(d)$ and $V(d)$ denote textual and visual features. The model yields both a score and a concise natural-language rationale. The score subsequently serves as rewards during the group relative policy optimization stage (Section~\ref{sec:grpo_stage}) to explicitly prioritize authoritative documents. Our validation also confirms that authority scores align strongly with human judgment (see Appendix~\ref{app:vlm_corr} for details).

\subsection{Three-Stage Training Pipeline}\label{sec:training_pipeline}

We systematically embed authority into retrievers with three distinct stages. For the generation target, we utilize host-level URLs as document identifiers\footnote{For instance, we use ``\texttt{plus.gov.kr}'' instead of ``\texttt{https://plus.gov.kr/portal/ntcmttr}''.}. The host-level granularity minimizes noise and exposes source identity, providing a stable foundation for authority modeling.

\subsubsection{Domain-Continued Pre-Training (CPT)}

The first stage adapts a general-purpose LLM to the search domain through domain-continued pre-training. This stage bridges the gap between broad linguistic knowledge and structured query–document correlations, as pointed out in the prior work~\cite{ye-etal-2025-best}\footnote{See Appendix~\ref{app:cpt} for further discussion.}. Specifically, we leverage large-scale search logs formatted as \texttt{[Query + URL + Title + Body]}. This structure enables the model to internalize associations between content and source identity. For instance, it learns that domains like ``\texttt{.gov}'' correlate with official institutions. Formally, the model parameters $\theta$ is updated via the standard language modeling objective:
\begin{equation}
\mathcal{L}_{\text{CPT}} = - \sum_{t} \log p_\theta(x_t \mid x_{<t}),
\end{equation}
where $x_t$ denotes the $t$-th token in the concatenated sequence. By treating URLs as meaningful semantic units rather than random strings, CPT establishes a robust prior for the subsequent supervised mapping and authority alignment stages.

\subsubsection{Supervised Fine-Tuning (SFT)}\label{sec:training_pipeline_sft}

The supervised fine-tuning stage optimizes the model for generate relevant DocIDs given a query. This step transforms the latent query-URL associations from CPT into a robust ranking capability. Formally, for a query-document pair $(q,d)$ from the dataset $\mathcal{D}$, we minimize the negative log-likelihood of the ground truth DocID sequence:
\begin{equation}
\label{eq:sft_loss}
\mathcal{L}_{\text{SFT}} = - \mathbb{E}_{(q,d) \sim \mathcal{D}} \big[ \log P_\theta(d \mid q) \big],
\end{equation}
\noindent
where $P_\theta(d \mid q) = \prod_{t=1}^{L} p_\theta(y_t \mid q, y_{<t})$ denotes the probability of generating the DocID sequence $y = (y_1,\ldots,y_L)$\footnote{Refer Appendix~\ref{app:sft_prompt} for the input prompt.}.

To construct scalable training data, we utilize real-world search click logs from a major commercial search engine. These logs provide continuously updated signals of user intent and emerging trends, yet inherently suffer from noise due to position bias and accidental clicks. To mitigate this, we employ a hybrid filtering strategy: (i) frequency-based pruning to remove unstable long-tail queries and (ii) relevance verification using an auxiliary ranker to discard semantically mismatched pairs. This approach effectively balances data scale with quality, establishing a reliable foundation for the subsequent authority alignment stage.

\subsubsection{Authority-Aware Ranking with Group Relative Policy Optimization (GRPO)}
\label{sec:grpo_stage}

\begin{figure}[t]
\centering
\includegraphics[width=0.95\linewidth]{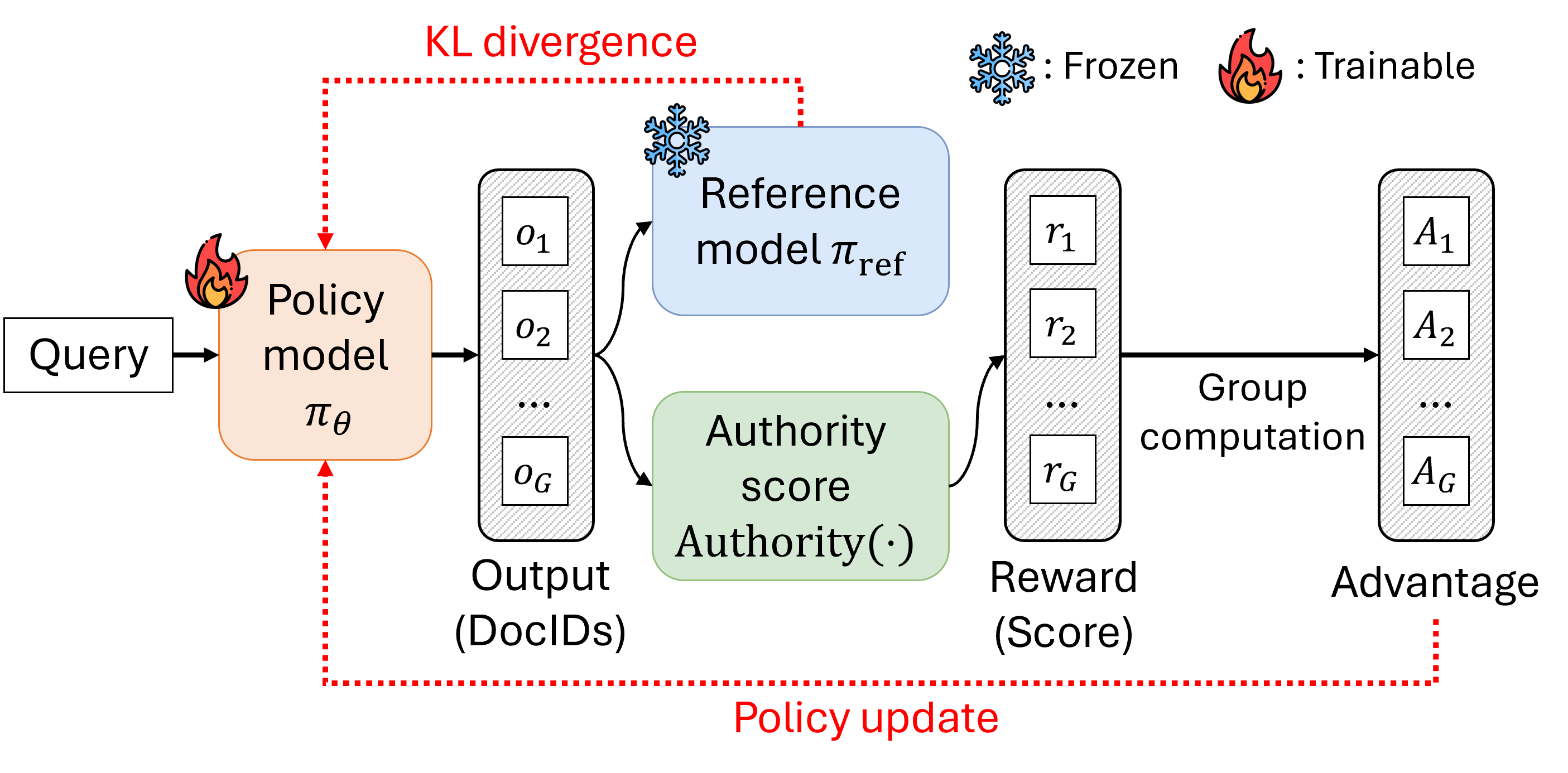}
\vspace{-2mm}
\caption{Illustration of the GRPO stage.}\label{fig:grpo}
\vspace{-2mm}
\end{figure}

The final stage employs preference optimization to align the model with authority signals. Since SFT treats all valid documents equally, it fails to capture the inherent relativity of trustworthiness. While simple alternatives like Weighted Cross-Entropy exist, they remain pointwise and lack the exploration mechanisms necessary for effective ranking\footnote{We further discuss the limitations in Appendix~\ref{sec:appendix_wce}.}. To overcome these limitations, we adopt Group Relative Policy Optimization (GRPO)~\cite{deepseek2024deepseek-math-grpo}, enabling the model to explicitly prioritize authoritative sources within a group.

As illustrated in Figure~\ref{fig:grpo}, we sample a group of $G$ candidate DocIDs $O=\{o_1, \dots, o_G\}$ for a given query $q$ using the current policy $\pi_{\theta_{\text{old}}}$. Each output $o_i$ receives a scalar reward $r_i = \texttt{Authority}(d_i)$. We then compute the advantage $A_i=\tfrac{r_i-\text{mean}(\mathbf{r})}{\text{std}(\mathbf{r})}$ by normalizing rewards within the group. The policy is optimized via the following objective:
\begin{equation}\small
\label{eq:grpo}
\begin{split}
\mathcal{L}_{\text{GRPO}}
&= \mathbb{E}\left[{q \sim P(Q),\, \{o_i\}_{i=1}^G \sim \pi_{\theta_{\text{old}}}(O|q)} \right] \\
&\frac{1}{G}\sum_{i=1}^{G} \Bigl[ \min ( \rho_i A_i,\; \mathrm{clip}(\rho_i,1-\epsilon,1+\epsilon) A_i) \\
&- \beta\, D_{\mathrm{KL}}(\pi_\theta \,\|\, \pi_{\text{ref}})\Bigr],
\end{split}
\end{equation}

\noindent
where $\rho_i = \tfrac{\pi_\theta(o_i|q)}{\pi_{\theta_{\text{old}}}(o_i|q)}$ is the likelihood ratio. $\pi_{\theta}$ and $\pi_{\text{ref}}$ are the policy and reference model initialized with the SFT model. The hyperparameters $\beta$ and $\epsilon$ regulate the KL penalty and the clipping range. This stage enables the model to distinguish authority levels beyond simple relevance by effectively balancing exploration and exploitation. For training stability, we utilize a  strictly filtered dataset of high-frequency queries using pre-computed rewards.

\subsection{Hybrid Ensemble Pipeline for Deployment}\label{sec:ensemble}
For production, we employ a Hybrid Ensemble Pipeline that synergizes existing retrievers with our model. This balances high recall and high precision in real-world scenarios. While existing retrievers~\cite{kwon-etal-2025-qupid} provides relevance scores $S_\text{rel}(d \mid q)$ for broad coverage, AuthGR generates a concise set of DocIDs $\mathcal{D}_\text{auth}(q)$ aligned with authority. For $d \in \mathcal{D}_\text{auth}(q)$, we derive a normalized score via linear decay:
\begin{equation}\small
S_{\mathrm{auth}}(d \mid q) = \frac{N - \texttt{rank}(d) + 1}{N}, 
\end{equation}
\noindent
where $N$ is the number of generated DocIDs. This formulation assigns higher weights to top-ranked candidates, effectively boosting authoritative documents in the final results.

The final score is computed as follows:

\begin{equation}\small
\label{eq:hybrid}
S_{\mathrm{final}}(d \mid q) = S_{\mathrm{rel}}(d \mid q) + \lambda \cdot S_{\mathrm{auth}}(d \mid q) \cdot \mathbb{I}\left[ d \in \mathcal{D}_{\mathrm{auth}}(q) \right]
\end{equation}
\noindent
where $\mathbb{I}[\cdot]$ is the indicator function and $\lambda$ controls the strength of authority signal. This formulation preserves the recall of existing rankers while injecting the knowledge of the AuthGR, yielding results that are both semantically relevant and trustworthy.

\section{Experimental Setup}

\noindent
\textbf{Datasets}.
We constructed three industrial-scale datasets from proprietary logs of a large-scale commercial web search engine in Korea. For CPT, we utilized 9.85M query–document pairs formatted as \texttt{[Query; URL; Title; Body]} from the web crawl of the search engine. For SFT, we curated 3.95M pairs from high-stakes domains such as health or finance with weekly query counts (QC) $>50$, filtering 63\% of noisy interactions. For GRPO, we selected 13.81K queries (QC $>200$) and utilized pre-computed authority scores across 3.75M host URLs as rewards. Please refer to Appendix~\ref{app:setup_data}.

\begin{table}[]\footnotesize
\centering
\setlength{\tabcolsep}{4pt} 
\renewcommand{\arraystretch}{1}

\begin{tabular}{lcccc}
\toprule
\multicolumn{1}{l|}{Model}        & \multicolumn{1}{c|}{Size} & P@3 & R@5 & R@10 \\ \midrule
\multicolumn{5}{l}{\textit{In-context Learning (ICL)}}                                                                                           \\ \midrule
\multicolumn{1}{l|}{Gemma 3}                & \multicolumn{1}{c|}{27B}                     & 0.1255               & 0.1732            & 0.2285             \\
\multicolumn{1}{l|}{EXAONE 3.5}                & \multicolumn{1}{c|}{32B}                     & 0.0825               & 0.1253            & 0.1612             \\
\multicolumn{1}{l|}{K-EXAONE}                & \multicolumn{1}{c|}{236B}                     & 0.1366               & 0.1918            & 0.2656             \\
\multicolumn{1}{l|}{Qwen3}                 & \multicolumn{1}{c|}{32B}                     & 0.0821               & 0.1176            & 0.1570             \\
\multicolumn{1}{l|}{LLaMA 3.1}             & \multicolumn{1}{c|}{405B}                    & 0.1413               & 0.1974            & 0.2590             \\
\multicolumn{1}{l|}{LLaMA 4 Scout}         & \multicolumn{1}{c|}{109B}                    & 0.1066               & 0.1555            & 0.2128             \\
\multicolumn{1}{l|}{LLaMA 4 Maverick}      & \multicolumn{1}{c|}{400B}                    & 0.1274               & 0.1841            & 0.2483             \\
\multicolumn{1}{l|}{DeepSeek-R1}           & \multicolumn{1}{c|}{671B}                    & 0.0891               & 0.1328            & 0.1718             \\
\multicolumn{1}{l|}{DeepSeek-V3}           & \multicolumn{1}{l|}{671B}                    & 0.1359               & 0.1932            & 0.2626             \\
\multicolumn{1}{l|}{DeepSeek-V3.2}           & \multicolumn{1}{l|}{685B}                    & 0.1398               & 0.2027            & 0.2729            \\
\multicolumn{1}{l|}{GPT-4o}                & \multicolumn{1}{c|}{-}                        & 0.1700               & 0.2348            & 0.3170             \\ \midrule
\multicolumn{5}{l}{\textit{Supervised Fine-tuning (SFT)}}                                                                                        \\ \midrule
\multicolumn{1}{l|}{HyperCLOVAX}           & \multicolumn{1}{c|}{0.5B}                    & 0.3470               & 0.4933            & 0.6634             \\
\multicolumn{1}{l|}{HyperCLOVAX}           & \multicolumn{1}{c|}{1.5B}                    & 0.3573               & 0.5058            & 0.6708             \\
\multicolumn{1}{l|}{LLaMA 3.2}              & \multicolumn{1}{c|}{1B}                      & 0.3479               & 0.4948            & 0.6679             \\
\multicolumn{1}{l|}{LLaMA 3.2}              & \multicolumn{1}{c|}{3B}                      & 0.3602               & 0.5108            & 0.6892             \\
\multicolumn{1}{l|}{T5Gemma 2}                & \multicolumn{1}{c|}{0.5B}                      & 0.3280               & 0.4646            & 0.6151             \\
\multicolumn{1}{l|}{T5Gemma 2}                & \multicolumn{1}{c|}{2B}                      & 0.3399               & 0.4805            & 0.6384             \\
\multicolumn{1}{l|}{Qwen3}                 & \multicolumn{1}{c|}{1.7B}                    & 0.3433               & 0.4853            & 0.6582             \\
\multicolumn{1}{l|}{Qwen3}                 & \multicolumn{1}{c|}{4B}                      & 0.3581               & 0.5053            & 0.6843             \\
\multicolumn{1}{l|}{HyperCLOVAX}           & \multicolumn{1}{c|}{14B}                     & 0.3854               & 0.5508            & 0.7289             \\ \midrule
\multicolumn{5}{l}{\textit{Ours}}                                                                                                                \\ \midrule
\multicolumn{1}{l|}{AuthGR (SFT)}          & \multicolumn{1}{c|}{3B}                      & 0.3555               & 0.5058            & 0.6899             \\
\multicolumn{1}{l|}{AuthGR (CPT+SFT)}      & \multicolumn{1}{c|}{3B}                      & 0.3725               & 0.5293            & 0.7031             \\
\rowcolor{gray!5} 
\multicolumn{1}{l|}{AuthGR (Full)} & \multicolumn{1}{c|}{3B}                      & \textbf{0.3856}      & \textbf{0.5464}   & \textbf{0.7175}    \\ 
\bottomrule

\end{tabular}
\vspace{-0.5em}
\caption{Offline evaluation results. Bold denotes the best performance of our method. `AuthGR (Full)' denotes the final model incorporating CPT, SFT, and GRPO.}\label{tab:offline_overall}
\vspace{-1.5em}

\end{table}

\noindent
\textbf{Baselines}.
We adopt two categories of baselines considering Korean as the target language. (i) In-context learning: Large-scale foundation models including \textbf{Gemma 3}~\cite{google-deepmind2025gemma}, \textbf{EXAONE 3.5}~\cite{lg2024exaone3.5}, \textbf{K-EXAONE}~\cite{lg2025k-exaone}, \textbf{Qwen3}~\cite{qwen-team2025Qwen3}, \textbf{LLaMA 3.1}~\cite{meta2024LLaMA3}, \textbf{LLaMA 4}~\cite{meta2025llama4}, \textbf{DeepSeek-R1/V3}~\cite{deepseek-ai2025deepseekr1,deepseek-ai2024deepseek0v3}, and \textbf{GPT-4o}~\cite{openai2024gpt-4o}. (ii) Supervised fine-tuning: Compact models including \textbf{HyperCLOVAX}~\cite{naver2024HyperClOVAX, team2025hyperclova}, \textbf{T5Gemma 2}~\cite{google2025T5Gemma2}, \textbf{LLaMA 3.2}~\cite{meta2024LLaMA3}, and \textbf{Qwen3}~\cite{qwen-team2025Qwen3}. See Appendix~\ref{app:setup_baseline} for details.

\noindent
\textbf{Implementation Details}.
We initialized a 3B decoder-only transformer from the language component of HyperCLOVAX-SEED-Vision-Instruct~\cite{naver2024HyperClOVAX}. For GRPO,  a rollout group size was $G=256$ and KL coefficient was $\beta=0.2$. During inference, we employed beam search with a size of 10, and the coefficient of the ensemble was set to $\lambda=0.6$ based on validation tuning. Further details are in Appendix~\ref{app:setup_detail}.

\noindent
\textbf{Evaluation Protocols}.
We validate AuthGR via three evaluation protocols. (i) \textbf{Offline Evaluation}: We utilize 3,000 expert queries with human-labeled ground truth to measure Precision@3 and Recall@\{5,10\}. (ii) \textbf{Human Evaluation}: A blind side-by-side comparison on 500 queries assesses the quality of AuthGR-enhanced pipeline over production system. (iii) \textbf{Online A/B Test}: We analyze user engagement metrics across millions of interactions on the large-scale web search platform. We compare the production system against AuthGR-enhanced pipeline, monitoring key metrics including click-through rate and total document clicks in mid-2025. See Appendix~\ref{app:setup_eval} for details.

\begin{table}[t]\footnotesize
\centering
\renewcommand{\arraystretch}{1.0}
\begin{tabular}{c|cc}
\toprule
                & Production  & Hybrid Ensemble \\
\midrule
Label score     & 3.06   & \textbf{3.41} \\
\bottomrule
\end{tabular}
\caption{Human evaluation results in a blind side-by-side test, comparing (i) the production system and (ii) its integration with AuthGR via Hybrid Ensemble.}\label{tab:human_eval}

\end{table}

\begin{table}[t]\footnotesize
\centering

\begin{tabular}{lcc}
\toprule
Metrics & Control & Treatment \\ \midrule
Pages with clicks   & +0.08\% & +21.36\%  \\
Total document clicks & +0.22\%  & +22.07\%    \\
Top 1 document CTR & +0.87\%  & +22.83\%    \\
Top 3 document CTR & +0.81\%  & +22.68\%    \\
Top 5 document CTR & +0.81\%  & +22.76\%    \\
\bottomrule
\end{tabular}
\vspace{-0.5em}

\caption{Online A/B test results showing relative gain in user engagement. `Control' group uses the production system. `Treatment' group receives results enhanced by AuthGR through the Hybrid Ensemble.}\label{tab:online_abtest}

\end{table}

\section{Experimental Results}\label{sec:results}

\subsection{Main Results}

\noindent
\textbf{Offline Evaluation}.
Table~\ref{tab:offline_overall} demonstrates the offline performance, where AuthGR achieves superior accuracy with the highest P@3 among all baselines. Remarkably, our 3B model performs on par with the 14B model despite being 4.7$\times$ smaller, highlighting the parameter efficiency of our framework. Moreover, the effectiveness of three-stage pipeline is validated, where CPT ensures domain adaptation and GRPO explicitly optimizes authority. Furthermore, the limited performance of ICL baselines underscores the necessity of task-specific fine-tuning for effective generative retrieval.

\noindent
\textbf{Human Evaluation}.
Table~\ref{tab:human_eval} presents the results of a blind side-by-side comparison between the commercial search engine's production system and AuthGR-enhanced ensemble pipeline. Each result was evaluated on a 1--5 point scale across relevance and authority. Remarkably, AuthGR achieves 11.4\% gain in average score, confirming that explicitly optimizing for authority yields results that users perceive as more relevant and trustworthy.

\noindent
\textbf{Online A/B Test}.
As reported in Table~\ref{tab:online_abtest}, AuthGR delivers substantial improvements in an online A/B test. Compared to the production baseline, user engagement metrics surged, where `Pages with clicks’ increased by 21.36\% and `Total document clicks’ by 22.07\%. Notably, the ‘Top-1 document CTR’ is enhanced by 22.83\%, indicating that authority-aware ranking significantly enhances the quality of the top-ranked results.

\subsection{In-depth Analysis}
\noindent
\textbf{Impact of Training Stages}.
Table~\ref{tab:ablation_stage} illustrates an ablation study for the training stages of AuthGR, highlighting four key insights. (i) The full pipeline achieves the best performance, yielding a total gain of 8.5\% in P@3 over the SFT baseline for the 3B model. (ii) The CPT stage delivers a dominant improvement of 4.8\%, creating a strong domain-specific foundation for the subsequent GRPO stage, which contributes an additional 3.0\%. (iii) The benefit of GRPO scales with model capacity. The 3B model exhibits a larger gain of 3.1\% compared to 1.3\% for the 0.5B model, suggesting that larger models better internalize the concept of authority. (iv) A clear synergy exists between stages, GRPO achieves a higher gain of 3.5\% when applied after CPT, compared to 3.0\% for the SFT-only baseline.

\begin{table}[t]\footnotesize
\vspace{-1.5mm}
\setlength{\tabcolsep}{3.6pt}
\renewcommand{\arraystretch}{1}
\centering

\begin{tabular}{c|clc|ccc}
\toprule
Model & SFT & \multicolumn{1}{c}{CPT} & GRPO & P@3 & R@5 & R@10 \\ \midrule
AuthGR & \checkmark &  & \multicolumn{1}{l|}{} & 0.3470 & 0.4933 & 0.6634 \\
 (0.5B)  & \checkmark &  & \checkmark & \textbf{0.3515} & \textbf{0.4989} & \textbf{0.6651} \\ \midrule
\multirow{4}{*}{\begin{tabular}[c]{@{}c@{}}AuthGR \\ (3B)\end{tabular}} & \checkmark &  & \multicolumn{1}{l|}{} & 0.3555 & 0.5058 & 0.6899 \\
 & \checkmark &  & \checkmark & 0.3660 & 0.5216 & 0.6986 \\
 & \checkmark & \multicolumn{1}{c}{\checkmark} & \multicolumn{1}{l|}{} & 0.3725 & 0.5293 & 0.7031 \\
 & \checkmark & \multicolumn{1}{c}{\checkmark} & \checkmark & \textbf{0.3856} & \textbf{0.5464} & \textbf{0.7175} \\ \bottomrule
\end{tabular}
\caption{Ablation study of the training stages.}\label{tab:ablation_stage}

\end{table}

\begin{figure}[t]\small
\centering
\includegraphics[width=0.95\linewidth]{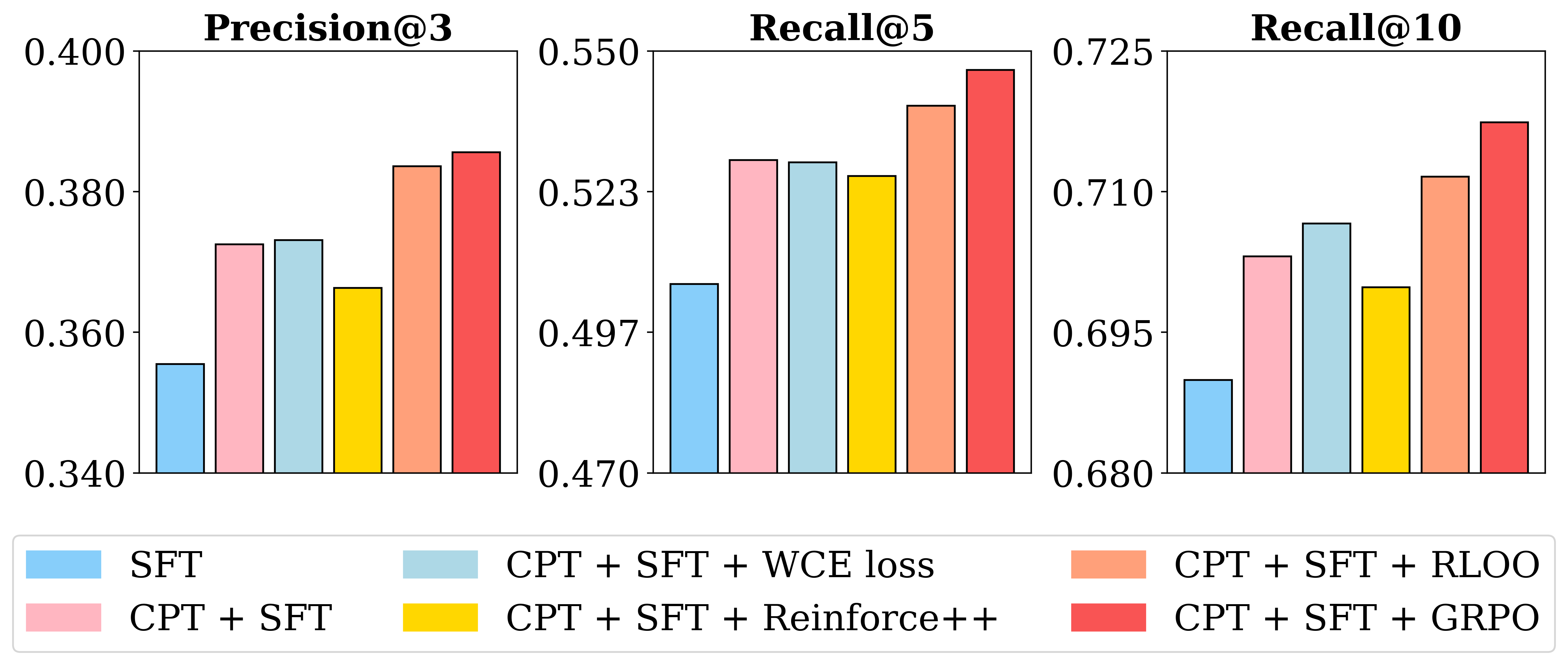}
\vspace{-0.5em}
\caption{Performance of ranking optimization methods. `CPT+SFT+GRPO' represents our AuthGR.}
\label{fig:ranking_algorithm}
\end{figure}

\noindent
\textbf{Comparison of Ranking Optimization}. 
Figure~\ref{fig:ranking_algorithm} compares the effectiveness of GRPO against representative pointwise methods such as Weighted Cross-Entropy (WCE) and Reinforce++~\cite{hu2025reinforce++}, as well as groupwise alternatives like RLOO~\cite{ahmadian2024RLOO}. The key findings are as follows: (i) GRPO emerges as the most effective approach, yielding the highest gain of 3.5\% in P@3 over the CPT + SFT baseline. (ii) Groupwise methods significantly outperform pointwise methods. GRPO and RLOO deliver clear gains of 3.5\% and 3.0\% in P@3, respectively. In contrast, WCE provides only a marginal 0.2\% gain, and Reinforce++ degrades performance by 1.7\%. This confirms that modeling the relative order of candidates is better suited for the ranking tasks.

\begin{table}[t]\footnotesize
\vspace{-1.5mm}
\setlength{\tabcolsep}{2pt}
\renewcommand{\arraystretch}{1}
\centering
\begin{tabular}{lcccc}
\toprule
\multirow{2}{*}{Training stage} & \multicolumn{2}{c}{Authority scores} & \multicolumn{2}{c}{\# docs by authority} \\
\cmidrule(lr){2-3} \cmidrule(lr){4-5}
& Mean($\uparrow$) & Median($\uparrow$) & Low($\downarrow$) & High($\uparrow$) \\
\midrule
CPT+SFT      & 87.2          & 90.0          & 118             & 2,754 \\
+ GRPO (Binary)           & 88.0          & 90.0          & 115   & 2,804 \\
+ GRPO (Linear) & \textbf{90.4} & \textbf{95.0} & \textbf{106} & \textbf{2,877} \\

\bottomrule
\end{tabular}%
\caption{Authority score distribution in generated documents. Mean and Median are computed on a 0-100 scale. `Low' and `High' indicate the number of documents scoring 0-60 and 90-100, respectively.}\label{tab:authority_distribution}
\vspace{-1em}
\end{table}

\noindent
\textbf{Impact on Authority Scores}.
Table~\ref{tab:authority_distribution} shows that GRPO successfully reshapes the authority distribution of generated documents. The mean score increases from 87.2 to 90.4, with the median reaching 95.0. This distributional shift is driven by a 10.2\% reduction in low-authority generations ($0$–$60$) and a 4.5\% increase in high-authority generations ($90$–$100$). This confirms that the model systematically learns to prioritize trustworthy sources.

\section{Conclusion}

We propose AuthGR, a generative retrieval framework that explicitly integrates document authority via multimodal scoring and progressive training. By employing a hybrid ensemble strategy, we successfully deployed this 3B model to a commercial search engine, where it matched the performance of 14B baselines and significantly improved real-world user engagement and satisfaction. Our work underscores the critical role of authority in generative retrieval, paving the way for the development of trustworthy real-world search applications.

\section{Limitations}\label{sec:limitation}
While AuthGR demonstrates significant improvements in authority-aware generative retrieval, we acknowledge several limitations that provide directions for future work. (i) We primarily use multimodal authority scores as reward signals for policy optimization. Although these scores prove the effectiveness in capturing site trustworthiness, incorporating more diverse and granular reward signals could further refine the model's understanding. For instance, integrating implicit user feedback, such as dwell time or scroll depth, could further enhance user satisfaction. (ii) We observed that authority reasoning capabilities scale with model size, as the 3B model significantly outperformed the 0.5B model. However, we did not scale beyond 3B due to the computational overhead. By leveraging advanced engineering techniques for inference efficiency, such as model quantization or speculative decoding, future research could explore the scalability of AuthGR with much larger foundation models to further unlock complex authority reasoning potential.

\section*{Ethics Statement}

This work fully complies with the ACL Ethics Policy. We declare that there are no ethical issues in this paper. The scientific artifacts we have utilized are publicly available for research under permissive licenses, and the utilization of these tools is consistent with their intended applications.

\bibliography{references}

\appendix

\newpage 
\newpage
\appendix

\section{Additional Details on Multimodal Authority Scoring}\label{app:vlm}

\subsection{VLM Prompt and Full Output Schema}\label{app:vlm_prompt}
As shown in Figure~\ref{fig:vlm_prompt}, the VLM prompt is designed with a systematic structure to ensure reliability and consistency. It comprises six key  components: \textit{Context} provides the background information for explaining the situation; \textit{Role} defines the expert persona; \textit{Task} specifies the concrete action the model is expected to perform; \textit{Output format} prescribes the structure of the response; \textit{Instruction} offers detailed guidance; and \textit{Constraints} outline the strict conditions. This structured design enables consistent authority evaluation across diverse websites. The VLM processes multimodal inputs including textual data and screenshots of websites. Multiple sites are grouped and processed through a batch API request. The results are returned as a strictly formatted JSON array, and \textbf{this predefined structure eliminates complex parsing overhead}, thereby significantly enhancing data processing efficiency. Note that unevaluable scores -1 are mapped to 0.

\begin{table}[t]\small
\centering
\vspace{-1.5mm}

\begin{tabular}{c|c}
\toprule
Metric & Value  \\ \midrule
Point-biserial correlation ($r$) & 0.495 ($p < 0.001$)   \\
ROC-AUC & 0.915  \\
Matthews Correlation (MCC) & 0.526\\ \bottomrule
\end{tabular}

\caption{Quantitative alignment between VLM-generated authority scores and human expertise labels.}\label{tab:auth_score_correlation}
\vspace{-0.5em}

\end{table}

\subsection{Reliability of VLM Scores}\label{app:vlm_corr}
To validate the VLM scores as a reliable proxy for human judgment, we analyzed their correlation with binary human expertise labels (true/false). As summarized in Table~\ref{tab:auth_score_correlation}, \textbf{all metrics consistently demonstrate strong and statistically significant correlation between VLM scores and human assessments}. The Point-biserial correlation and Matthews Correlation Coefficient confirm robust positive alignment between the VLM scores and the expert-labeled expertise. Additionally, the ROC-AUC demonstrates the discriminative power of the model in distinguishing authoritative sites. These results verify that the VLM effectively functions as a high-quality automated assessor that closely reflects human evaluators' intuition.

\begin{figure}[t]
\centering
\vspace{-5.5mm}
\includegraphics[width=0.9\linewidth]{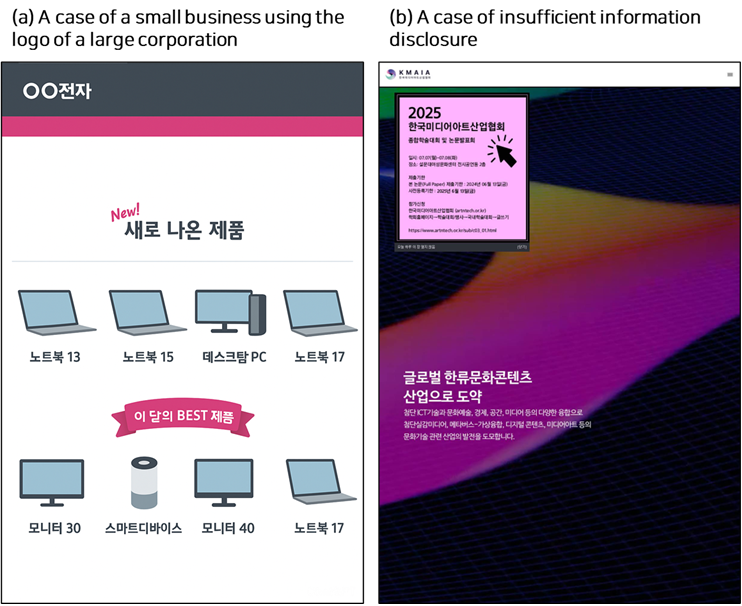}
\caption{Examples where the Text+Image input yields more appropriate results than the image-only input.}
\label{fig:vlm_modality_examples}
\end{figure}

\subsection{Impact of Modalities on Scoring}\label{app:vlm_modality}
To assess the contribution of different modalities, we evaluated the VLM on 40,000 websites using three settings: text-only (URL, title, body, Wikipedia), image-only (screenshots), and text+image. Each website was independently annotated by human annotators, who labeled whether the site exhibited expertise (true or false). We then considered a site to be authoritative if the VLM produced a high authority score, and measured agreement with the human judgments. The VLM achieved approximately 81\% accuracy with text-only inputs and 92\% with image-only inputs. Notably, \textbf{the multimodal setting reached 97\% accuracy, demonstrating that multimodal inputs substantially improve authority detection}.

\begin{figure*}[!htbp]\footnotesize
\centering
\begin{tcolorbox}[
  enhanced,
  colframe=black,
  colback=white!5!white,
  title=\textbf{VLM Prompt},
  fonttitle=\bfseries
]
\footnotesize
\textbf{Context:}\\
The task is to determine which web documents are valuable enough to be included in search engine results. The goal is to provide users with high-quality web documents that demonstrate reliability and expertise. Such documents typically originate from authoritative institutions or organizations, or consist of content created with significant effort and originality.

\vspace{2mm}
\footnotesize
\textbf{Role:}\\
You are a site authority evaluation expert.

\vspace{2mm}
\textbf{Task:}
\begin{enumerate}[leftmargin=*, nosep]
  \item \textbf{Site type:} Classify the website by its structure/function.
  \item \textbf{Topic:} Categorize the main subject or information the site covers.
  \item \textbf{Score:} Assign an integer score between -1 and 100.
  \item \textbf{Evidence:} Provide 1--2 sentences in explaining the reason for the score.
\end{enumerate}

\vspace{2mm}
\textbf{Output Format:}\\
All output must be a JSON Array as shown below:

\begin{tcblisting}{
    listing only,
    colback=gray!8,
    colframe=gray!40,
    boxrule=0.2pt,
    arc=0.8mm,
    left=1.5mm, right=1.5mm, top=1mm, bottom=1mm,
    listing options={
        basicstyle=\ttfamily\scriptsize,
        breaklines=true,
        breakindent=0pt,
        columns=fullflexible
    }
}
[{"url": "", "site_type": "", "topic": "", 
"score": 0, "evidence": ""}, ...]
\end{tcblisting}

\begin{itemize}[leftmargin=*, nosep]
  \item Do not include any text outside of the JSON Array.
  \item The number of output items must match the number of input sites.
\end{itemize}

\vspace{2mm}
\textbf{Site Type Instruction:}
\begin{itemize}[leftmargin=*, nosep]
  \item \textbf{Commerce}: A site where users can directly purchase products, with visible product listings showing prices.
  \item \textbf{SNS}: An official website where social interaction and content sharing are the main features.
  \item \textbf{Community}: An official site for discussions and information exchange with forums, comments, or membership.
  \item \textbf{News/Media}: An official news or broadcasting site run by governments, newspapers, or broadcasters.
  \item \textbf{Blog}: Blog platforms.
  \item \textbf{General}: Any site not fitting the above categories, or where the type is ambiguous.
\end{itemize}

\vspace{2mm}
\textbf{Topic Instruction:}
\begin{itemize}[leftmargin=*, nosep]
  \item \textbf{Health}: Medical info, diseases, healthcare, medicine, devices, nutrition, rehab, diet.
  \item \textbf{Education}: Schools, learning, careers, exams, certificates, training, textbooks, courses.
  \item \textbf{IT}: Tech, telecom, electronics, digital devices, industry news, trends.
  \item ...
\end{itemize}

\vspace{2mm}
\textbf{Score Instruction:}
\begin{itemize}[leftmargin=*, nosep]
  \item 80--100: High-quality sites
  \item 50--79: Mid-quality sites
  \item 10--49: Low-quality sites
  \item 0: Spam, gambling, illegal, adult content, \texttt{warning.or.kr}, or policy-violating content
  \item -1: No information, inaccessible site, or unable to evaluate
\end{itemize}

\vspace{2mm}
\textbf{Constraints:}
\begin{itemize}[leftmargin=*, nosep]
  \item Strictly adhere to the JSON output format.
  \item The \texttt{evidence} field must always be written in Korean.
  \item Prioritize the screenshot image over other signals when evaluating.
  \item Do not classify \texttt{site\_type} solely by URL.
\end{itemize}

\end{tcolorbox}
\caption{Prompts used for Multimodal Authority Scoring.}\label{fig:vlm_prompt}

\end{figure*}

\subsection{Efficiency and Scalability of Scoring}\label{app:vlm_cost}
\noindent
\textbf{Cost Efficiency}. We adopt three optimization strategies. First, we employ \textbf{batch requests}, where multiple queries are grouped and sent to the API simultaneously. This reduces the number of calls, minimizes network overhead, and benefits from lower per-request costs compared to real-time queries. Second, \textbf{multi-sample requests} are utilized by consolidating multiple sites into a single prompt, thereby reducing prompt token usage and improving overall response efficiency. Finally, we enforce \textbf{structured JSON output formats} instead of free-form natural language. This eliminates unnecessary descriptive text, lowers post-processing overhead, and enables seamless integration into automated pipelines. Together, these strategies significantly enhance the scalability and economic viability of VLM-based authority evaluation. 

\noindent
\textbf{Scalability and Maintenance}.
For production-scale deployment, we conduct a full database re-scoring quarterly, requiring approximately 30M tokens per 100K sites. While we used a high-capacity VLM internally, we confirmed that open-source VLMs effectively reproduce our results. This validates generalizability of the rubric beyond proprietary models. To ensure freshness, we monitor content changes during crawling and perform incremental weekly updates. We identify target sites using a Gradient Boosted Regression Tree model trained on VLM labels; sites whose predicted scores deviate significantly from current values are flagged for re-scoring. Combined with manageable retraining cost, this two-tier update strategy ensures sustainable real-world deployment.

\subsection{Qualitative Examples of Scoring}\label{app:vlm_example}

Relying solely on visual cues can lead to misjudgments when screenshots are misleading or insufficient.  Figure~\ref{fig:vlm_modality_examples}(a) illustrates a small vendor website selling products from a large corporation may display the corporation’s logo on its site. The image-only model may mistakenly identify it as the corporation’s official website. The VLM produced the incorrect evidence, "\textit{the official OO Electronics website — a corporate, authoritative site}," and consequently judged the site to exhibit high expertise. In contrast, in the text+image condition, the VLM correctly recognizes it as "\textit{the homepage of an individual or small-scale seller of OO Electronics products.}", assigning appropriate low expertise.

Similarly, Figure~\ref{fig:vlm_modality_examples}(b) shows a site with sparse visual content. The image-only model may dismiss it as  assessed low authority with the reasoning "\textit{the information site of the Art and Science Convergence Project Group, but the content is limited to a brief notice page.}" In contrast, by leveraging both texts and images, the model correctly identifies it as "\textit{official site of the Korea Media Art Industry Association, providing industry information and networking opportunities}," yielding a high score.

\begin{table}[t]\footnotesize
\centering
\setlength{\tabcolsep}{3pt}
\begin{tabular}{l|c|cc}
\toprule
Model & Size & Latency (ms) & Throughput \\ \midrule
HyperCLOVAX (SFT) & 14B & 2,881 & 1.30 \\
\textbf{AuthGR (Ours)} & \textbf{3B} & \textbf{1,225} & \textbf{3.26} \\ \bottomrule
\end{tabular}
\vspace{-2.5mm}
\caption{Comparison of inference efficiency. Latency denotes the average response time, and throughput indicates the number of requests processed per second.}\label{tab:efficiency_infer}
\end{table}

\section{Efficiency Analysis}
\subsection{Training Efficiency}
The three-stage pipeline is designed for periodic retraining at a manageable cost. On 8$\times$A100 GPUs, the full process completes in approximately 83 hours: (i) CPT: 57h (9.85M samples), (ii) SFT: 20h (3.95M samples), and (iii) GRPO: 6h (13.8K samples). The minimal overhead of the GRPO stage is particularly advantageous for frequent authority-alignment updates.

\subsection{Inference Efficiency}
Table~\ref{tab:efficiency_infer} compares the inference latency of AuthGR (3B) against the HyperCLOVAX-14B (SFT) baseline. On an NVIDIA A100 environment, AuthGR 3B achieves a 2.35$\times$ reduction in latency and 2.51$\times$ higher throughput while maintaining comparable ranking performance. This demonstrates that AuthGR delivers high-quality retrieval at a fraction of the computational cost of larger models.

\section{Extended Related Work on GenIR}\label{app:related_work}
\noindent
\textbf{DocID Design and Variations.} Beyond basic numeric IDs~\cite{nips/WangHWMWCXCZL0022/NCI, corr/abs-2206-10128/DSI-QG} and URLs~\cite{corr/abs-2208-09257/Ultron, cikm/ChenZG0FC22/CorpusBrain}, various DocIDs have been explored to bridge the gap between identifiers and document semantics. These include N-grams for substring-based retrieval~\cite{iclr/CaoI0P21/GENRE, nips/BevilacquaOLY0P22/SEAL, sigir23UGR}, or hierarchical codebooks~\cite{nips23MEVI, www/Zeng24RIPOR, sigir/Zeng24PAG}. 

\noindent
\textbf{Optimization and Learning Paradigms.} The evolution of GenIR has shifted from simple tasks to complex ranking optimization. Apart from the ranking losses~\cite{nips/23GENRET}, distillation have been employed to transfer ranking capabilities from large-scale teacher models to efficient generative retrievers~\cite{acl24DGR, www/Zeng24RIPOR}. Furthermore, the generative paradigm has been successfully adapted to recommendation systems, where models generate item IDs for personalized discovery~\cite{Rajput23TIGER, Geng0FGZ22P5, ZhengHLCZCW24LCRec, lee2025GRAM, emnlp2025grut, cikm/Ju25GRID, Kuaishou2025OneRec, corr/abs-2509-13648GenPAS}.

\noindent
\textbf{Applications and Feedback Signals.} Recent research incorporates diverse feedback signals, such as user preference rankings, to fine-tune generative models through reinforcement learning~\cite{eacl/Song24Re3val}. While these methods have been rapidly expanded into real-world industrial applications~\cite{recsys/Penha24Spotify, Google2025PLUM}, such as financial services~\cite{sigir24AlipayLLMGR}, visual discovery~\cite{corr/abs-2504-10507PinRec} or advertisement~\cite{sigir/Xi25LASER} platforms, they remain limited to relevance-centric optimization. Our work extends this trajectory by incorporating authority signals into the GenIR pipeline.

\section{Additional Experimental Setup}\label{app:setup}

\subsection{Datasets}\label{app:setup_data}

\begin{itemize}[leftmargin=*,topsep=0pt,itemsep=-1ex,partopsep=1ex,parsep=1ex]
\item \textbf{CPT Phase}: To adapt the model for generative retrieval, we designed diverse sequence formats. In addition to standard search log sequences such as \texttt{[Query; Title]} and \texttt{[Query; Snippet]}, we added \texttt{[Query; URL; Title; Body]}. This enables the model to learn the structural relationship between URLs and content reliability while modeling query-document relevance.
\item \textbf{SFT Phase}: We applied a two-stage refinement: (i) rule-based filtering to exclude low-authority platforms (\eg, personal blogs, inactive wikis) and (ii) model-based cleaning using a retriever~\cite{kwon-etal-2025-qupid}. Although this removed 63\% of raw click logs, our preliminary study showed that training on the curated subset improved P@3 by +16.62\% and accelerated convergence by 40\% compared to using the full, noisy dataset.
\item \textbf{GRPO Phase}: We focused on high-frequency queries across eight high-stakes domains: \textit{health, education, information technology, finance, parenting, society, animals,} and \textit{recruitment}. From weekly logs, we sampled queries with QC $> 50$ and further filtered for unambiguous user intents using an LLM (\texttt{gemma-3-27b-it}), resulting in 13,814 high-quality samples to ensure clear reward signals during policy optimization.

\end{itemize}

\subsection{Baselines}\label{app:setup_baseline}
All baselines were sourced from the Hugging Face Hub. The specific versions are listed below.

\noindent \textbf{(i) In-context Learning}: 
\begin{itemize}[leftmargin=*, topsep=2pt, itemsep=1pt, parsep=0pt] 
\item \textbf{Gemma 3}: \raggedright \href{https://huggingface.co/google/gemma-3-27b-it}{\normalfont\nolinkurl{google/gemma-3-27b-it}} 
\item \textbf{EXAONE 3.5}: \raggedright \href{https://huggingface.co/LGAI-EXAONE/EXAONE-3.5-32B-Instruct}{\normalfont\nolinkurl{LGAI-EXAONE/EXAONE-3.5-32B-Instruct}} 
\item \textbf{K-EXAONE}: \raggedright \href{https://huggingface.co/LGAI-EXAONE/K-EXAONE-236B-A23B}{\normalfont\nolinkurl{LGAI-EXAONE/K-EXAONE-236B-A23B}} 
\item \textbf{Qwen3}: \raggedright \href{https://huggingface.co/Qwen/Qwen3-32B}{\normalfont\nolinkurl{Qwen/Qwen3-32B}} 
\item \textbf{LLaMA 3.1}: \raggedright \href{https://huggingface.co/meta-llama/Llama-3.1-405B}{\normalfont\nolinkurl{meta-llama/Llama-3.1-405B}} 
\item \textbf{LLaMA 4}: \raggedright \href{https://huggingface.co/meta-llama/Llama-4-Scout-17B-16E-Instruct}{\normalfont\nolinkurl{meta-llama/Llama-4-Scout-17B-16E-Instruct}}, \href{https://huggingface.co/meta-llama/Llama-4-Maverick-17B-128E-Instruct}{\normalfont\nolinkurl{meta-llama/Llama-4-Maverick-17B-128E-Instruct}} 
\item \textbf{DeepSeek-R1}: \raggedright \href{https://huggingface.co/deepseek-ai/DeepSeek-R1}{\normalfont\nolinkurl{deepseek-ai/DeepSeek-R1}} 
\item \textbf{DeepSeek-V3}: \raggedright \href{https://huggingface.co/deepseek-ai/DeepSeek-V3}{\normalfont\nolinkurl{deepseek-ai/DeepSeek-V3}} 
\item \textbf{DeepSeek-V3.2}: \raggedright \href{https://huggingface.co/deepseek-ai/DeepSeek-V3.2}{\normalfont\nolinkurl{deepseek-ai/DeepSeek-V3.2}} 
\end{itemize}

\noindent \textbf{(ii) Supervised Fine-tuning}: 
\begin{itemize}[leftmargin=*, topsep=2pt, itemsep=1pt, parsep=0pt] 
\item \textbf{HyperCLOVAX}: \raggedright \href{https://huggingface.co/naver-hyperclovax/HyperCLOVAX-SEED-Text-Instruct-0.5B} {\normalfont\nolinkurl{naver-hyperclovax/HyperCLOVAX-SEED-Text-Instruct-0.5B}}, \href{https://huggingface.co/naver-hyperclovax/HyperCLOVAX-SEED-Text-Instruct-1.5B} {\normalfont\nolinkurl{naver-hyperclovax/HyperCLOVAX-SEED-Text-Instruct-1.5B}}, \href{https://huggingface.co/naver-hyperclovax/HyperCLOVAX-SEED-Think-14B} {\normalfont\nolinkurl{naver-hyperclovax/HyperCLOVAX-SEED-Think-14B}} 
\item \textbf{LLaMA 3.2}: \raggedright \href{https://huggingface.co/meta-llama/Llama-3.2-1B-Instruct}{\normalfont\nolinkurl{meta-llama/Llama-3.2-1B-Instruct}}, \href{https://huggingface.co/meta-llama/Llama-3.2-3B-Instruct}{\normalfont\nolinkurl{meta-llama/Llama-3.2-3B-Instruct}} 
\item \textbf{T5Gemma 2}: \raggedright \href{https://huggingface.co/google/t5gemma-2-270m-270m}{\normalfont\nolinkurl{google/t5gemma-2-270m-270m}}, \href{https://huggingface.co/google/t5gemma-2-1b-1b}{\normalfont\nolinkurl{google/t5gemma-2-1b-1b}} 
\item \textbf{Qwen3}: \raggedright \href{https://huggingface.co/Qwen/Qwen3-1.7B}{\normalfont\nolinkurl{Qwen/Qwen3-1.7B}}, \href{https://huggingface.co/Qwen/Qwen3-4B}{\normalfont\nolinkurl{Qwen/Qwen3-4B}} 
\end{itemize}

\subsection{Prompts for SFT Stage}\label{app:sft_prompt}
Figure~\ref{fig:sft_prompt} shows the instruction prompt used for the SFT stage. For fair comparison, we utilize the identical prompt for all SFT baselines in Table~\ref{tab:offline_overall}.

\begin{figure}[t]\footnotesize
\centering
\begin{tcolorbox}[colframe=black,colback=white!10!white,title=\textbf{SFT Instruction Prompt},fonttitle=\bfseries]
\footnotesize
\textbf{Instruction:}  \\
You are a model that generates authoritative websites. Based on the user query, generate the most relevant and authoritative website URL. 

\vspace{2mm}
\noindent \textbf{Conditions:}
\begin{enumerate}[leftmargin=*, nosep]
    \item The website must be a trustworthy domain (e.g., .gov, .edu, official institutions, etc.).
    \item Generate a website that contains appropriate information to fulfill the user's request.
    \item The output should only include the website.
    \item Even if the input is incomplete, analyze the user's intent to generate the most suitable website.
    \item If no suitable website is found, return an empty output.
\end{enumerate}

\vspace{1mm}
\noindent \textbf{Input:} 
\newline - Query: \texttt{\{query\}} \\

\textbf{Output:} 
\newline - Site: \texttt{[Website URL ID]}

\end{tcolorbox}
\caption{Prompts used for the SFT stage. The \texttt{\{query\}} field is a placeholder for the user query.}\label{fig:sft_prompt}
\end{figure}

\subsection{Implementation Details}\label{app:setup_detail}

We implemented the model in PyTorch and trained it on 8 NVIDIA A100 GPUs using DeepSpeed ZeRO-3. We utilized AdamW optimizer with $\beta_{1}=0.9$, $\beta_{2}=0.95$, weight decay $=0.01$, a cosine scheduler and a warmup ratio of $0.03$. The maximum input length was 2,048 tokens. For the CPT stage, we used a global batch size of 256 and peak learning rate 5.0$\times 10^{-6}$. In the SFT stage, the model was trained for 3 epochs with a global batch size of 512 and a learning rate of $1\times 10^{-6}$. For the GRPO stage, the model was trained for one epoch. During the rollout, we sampled candidate DocIDs using a temperature of $1.5$, a top-$p$ of $0.8$, and a top-$k$ of $50$. For the WCE baseline (Equation~\ref{eq:wce_weight}), we set $\alpha=4.0$. During inference, we used beam search following existing works~\cite{nips/Tay/DSI, nips/WangHWMWCXCZL0022/NCI} with a beam size of $10$. The maximum generation length per DocID was 50 tokens, sufficient to cover host-level URLs.

\subsection{Evaluation Protocol}\label{app:setup_eval}

\noindent
\textbf{Offline Quantitative Evaluation}. For the 3,000 expert queries, we retrieved 30 candidate documents per query and instructed human annotators to identify the top-3 relevant ones. To ensure fair comparison, both ground truth labels and model predictions were evaluated at the host-URL level.

\noindent
\textbf{Human Evaluation}. In a blind side-by-side comparison, annotators evaluated the top-1 results from the production baseline and our hybrid system. We used a 5-point scale that simultaneously weights semantic relevance (providing sufficient information) and authority (originating from credible sources). For each query, annotators made a direct preference judgment (``\textit{Which result is better overall?}'') to compute the win ratio. Detailed annotation guidelines are in Appendix~\ref{app:human_eval}.

\noindent
\textbf{Online A/B Testing}. We conducted an A/B test over several days in mid-2025. The control group received results from the existing production system, while the treatment group received results from the same system augmented with AuthGR through the hybrid ensemble pipeline. We monitored three engagement metrics: (i) \textit{pages with clicks} (the number of pages receiving at least one click), (ii) \textit{total document clicks} (the total number of document clicks), and (iii) \textit{top@k document click-through rate (CTR)} for $k \in \{1,3,5\}$ to specifically quantify user engagement with high-ranked authoritative documents. 

\subsection{Details of the Weighted Cross-Entropy Baseline}\label{sec:appendix_wce}

To incorporate authority signals, we examine a pointwise Weighted Cross-Entropy (WCE), as shown in Figure~\ref{fig:ranking_algorithm}. The core idea is to scale the standard cross-entropy loss for each DocID by its authority score, thereby encouraging the model to assign higher probabilities to more authoritative sources. Formally, given a document $d$, its authority score is transformed into a weight:
\begin{equation}
\label{eq:wce_weight}
w_d = 1 + \alpha \cdot \frac{\texttt{Authority}(d)}{100}
\end{equation}
\noindent
where $\alpha$ is a hyperparameter that controls the weight of authority. The objective minimizes the weighted negative log-likelihood over the training data $\mathcal{D}$:
\begin{equation}
\mathcal{L}_{\text{WCE}} = - \mathbb{E}_{(q,d) \sim \mathcal{D}} \big[ w_d \log P_\theta(d \mid q) \big]
\end{equation}
However, WCE has critical limitations: (i) pointwise loss misaligns with the ranking tasks, and (ii) it lacks exploration capability. These limitations motivated our adoption of the group-wise GRPO.

\section{Additional Experimental Results}\label{app:exp_result}

\begin{table}[t]\small
\centering
\vspace{-1.5mm}

\begin{tabular}{c|c|cc}
\toprule
Metric & Gain & $p$-value & 95\% CI \\ \midrule
P@3 & 2.31\% & 0.0277 & {[}0.0012, 0.0158{]} \\
R@5 & 1.65\% & 0.0483 & {[}-0.0002, 0.0170{]} \\ 
\bottomrule
\end{tabular}

\caption{Statistical significance between AuthGR and HyperCLOVAX 14B. Gain represents the relative difference in metrics between AuthGR and the 14B model.}\label{tab:stat_baseline}

\end{table}

\begin{table}[t]\small
\centering
\vspace{-1.5mm}
\setlength{\tabcolsep}{3.6pt}

\begin{tabular}{l|c|cc}
\toprule
Training stages & Gain & $p$-value & 95\% CI \\ \midrule
SFT $\to$ CPT+SFT & +4.38\% & \textless 0.0001 & {[}0.0082, 0.0221{]} \\
CPT+SFT $\to$ Full & +3.00\% & \textless 0.0001 & {[}0.0062, 0.0155{]} \\
SFT $\to$ Full & +7.51\% & \textless 0.0001 & {[}0.0189, 0.0334{]} \\ 
\bottomrule
\end{tabular}

\caption{Statistical significance of the three-stage training. Gain denotes the relative improvement in metric values achieved by each incremental stage on P@3.}\label{tab:stat_stage}
\vspace{-0.5em}

\end{table}

\begin{table}[t]\footnotesize
\centering
\setlength{\tabcolsep}{6pt}
\renewcommand{\arraystretch}{0.9}
\begin{tabular}{c|ccc}
\toprule
Scaling strategies & P@3 & R@5 & R@10 \\ \midrule
Binary & 0.3820 & 0.5375 & 0.7133 \\
Binning & 0.3841 & 0.5420 & 0.7151 \\
Sigmoid & 0.3833 & 0.5377 & 0.7147 \\ \midrule
\textbf{Linear (AuthGR)} & \textbf{0.3856} & \textbf{0.5464} & \textbf{0.7175} \\ \bottomrule
\end{tabular}
\caption{Performance comparison over various scaling strategies for authority scores.}\label{tab:score_scaling}
\end{table}

\subsection{Statistical Significance Analysis}
To ensure that performance gains are robust, we conducted statistical validation on 3,041 queries using paired t-tests with bootstrap resampling (5,000 iterations). We also report the 95\% Confidence Interval (CI) to estimate the range of the true mean performance difference.

\noindent
\textbf{Comparison with 14B Baseline}.
Table~\ref{tab:stat_baseline} shows that AuthGR 3B significantly outperforms the 14B baseline in top-rank metrics including P@3 and R@5 with $p < 0.05$. Despite its smaller size, our model achieves statistically superior precision. This accuracy, combined with lower operational costs, underscores the practical suitability of AuthGR for production-scale deployment.

\noindent
\textbf{Effectiveness of Training Pipeline}.
Table~\ref{tab:stat_stage} demonstrates that every stage yields statistically significant gains. Notably, the full pipeline achieves a +7.51\% relative improvement in P@3 over the SFT baseline, validating that our progressive training approach effectively enhances authority-aware reasoning beyond standard supervised fine-tuning.

\subsection{Impact of Reward Scaling Strategies}

To effectively leverage our fine-grained signals, we evaluate four reward scaling strategies in GRPO: (i) \textit{Binary}~\cite{deepseek2024deepseek-math-grpo}, mapping scores above a fixed threshold (e.g., 80) to 1 and others to 0; (ii) \textit{Binning}, discretizing scores into five ordinal classes; (iii) \textit{Sigmoid}, nonlinear transformation to emphasize higher scores; and (iv) \textit{Linear}, which directly uses the raw scores. Table~\ref{tab:score_scaling} demonstrates that the Linear strategy consistently outperforms all alternatives across all metrics. This suggests that preserving fine-grained authority differences is more effective than coarse approximations.

\begin{figure}[t]\small
\centering
\includegraphics[width=1.0\linewidth]{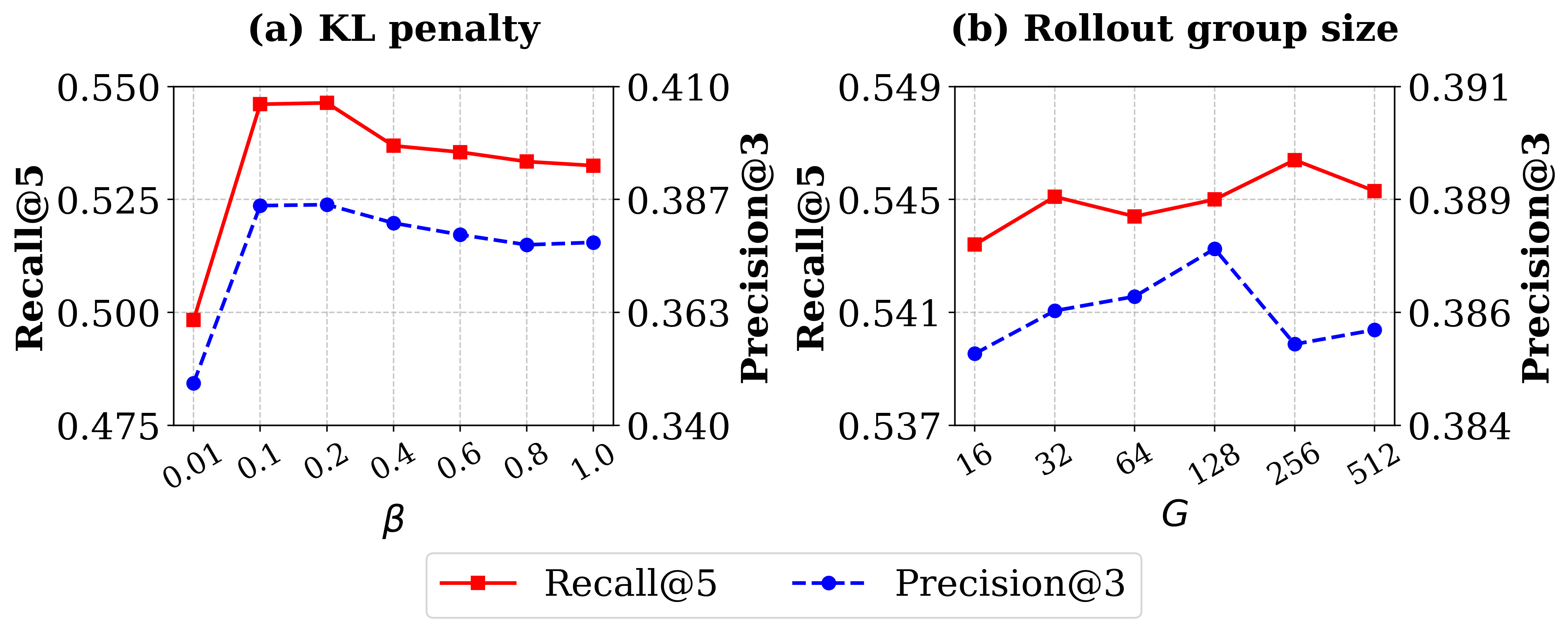}
\caption{Performance of AuthGR under different hyperparameters: (a) varying coefficient of the KL penalty $\beta$ and (b) varying size of rollout group $G$.}\label{fig:exp_hyper_kl}
\end{figure}

\subsection{Hyperparameter Study}
Figure~\ref{fig:exp_hyper_kl} shows the performance of AuthGR when varying the coefficient of the KL penalty $\beta$ and varying the size of the rollout group $n$. (i) The optimal $\beta$ lies in the range of 0.1-0.2. When setting $\beta$ as 0.2 from 0.01, the performance improves by 10.6\% and 9.7\% in Precision@3 and Recall@5, respectively. It highlights that model performance is highly sensitive to $\beta$. Moreover, explicitly regularizing the KL divergence between the trained and reference policies is crucial for preserving the relevance-based ID generation ability acquired during the SFT stage. (ii) Regarding rollout size, the best performance is observed when $n$ ranges from 128 to 512. In contrast, too few rollouts provide insufficient information to compute meaningful relative reward advantages within each group.

\begin{table}[t]
\centering

\resizebox{\linewidth}{!}{%
\begin{tabular}{lccc}
\toprule
Training stage & Recall & Avg. rank ($\downarrow$) & Avg. confidence ($\uparrow$) \\
\midrule
CPT+SFT (Baseline)      & 7,298             & 3.9246          & 0.2263 \\
+ GRPO (Binary)           & 7,438 (+1.9\%)    & \textbf{3.9091} & 0.2279 \\
+ GRPO (Linear) & \textbf{7,817 (+7.1\%)} & 3.9478 & \textbf{0.2288} \\
\bottomrule
\end{tabular}%
}
\caption{Impact of GRPO on accuracy and confidence. Recall counts queries with ground-truth DocID in the outputs. Avg. rank and confidence measure the position of ground-truth DocIDs and generation probability.}\label{tab:confidence}

\end{table}

\subsection{Impact on Retrieval Accuracy and Confidence}
To analyze the impact of GRPO on authority and confidence, Table~\ref{tab:confidence} compares the CPT+SFT baseline against GRPO-Binary (mapping scores $>$ 80 to 1) and GRPO-Linear (using raw scores). The Linear formulation proves most effective, achieving 7.1\% gain in Recall over the baseline and increased average confidence. This confirms that fine-grained linear rewards provide superior calibration compared to coarse binary signals, enabling the model to retrieve ground-truth documents with greater confidence.

\begin{table}[t]\footnotesize
\centering
\setlength{\tabcolsep}{6pt}
\renewcommand{\arraystretch}{0.9}
\begin{tabular}{c|ccc}
\toprule
Data filtering & P@3 & R@5 & R@10 \\ \midrule
w/o filtering & 0.3194 & 0.4642 & 0.6299 \\
w/ filtering & 0.3725 & 0.5293 & 0.7031 \\ \midrule
Gain (\%) & 16.62 & 14.02 & 11.62 \\
\bottomrule
\end{tabular}
\caption{Performance comparison with and without data filtering in the SFT stage. “Gain” represents the relative percentage improvement achieved after filtering.}\label{tab:data_filtering}
\end{table}

\subsection{Impact of Data Filtering in SFT}\label{app:exp_result_filtering}

Since raw click logs are inherently noisy, we investigated the impact of data quality on the SFT stage as shown in Table~\ref{tab:data_filtering}. Empirically, the best performance was obtained at a threshold of 2.1. At this point, approximately 63\% of the original data was filtered out, reducing the number of training samples from nearly 11 million to just over 4 million. Despite this substantial reduction, model performance improved markedly across all metrics with Precision@3 rising by 16.6\% and Recall@5 by 14.0\%, and Recall@10 by 11.6\%. These results provide strong empirical evidence that the \textit{quality} of supervision data is more critical than its \textit{quantity} in generative retrieval. By removing noise, the model can learn robust mappings between queries and documents.

\subsection{Performance over Data Filtering Thresholds}\label{app:exp_result_filtering_threshold}

\begin{figure}[t]
\centering
\includegraphics[width=0.8\linewidth]{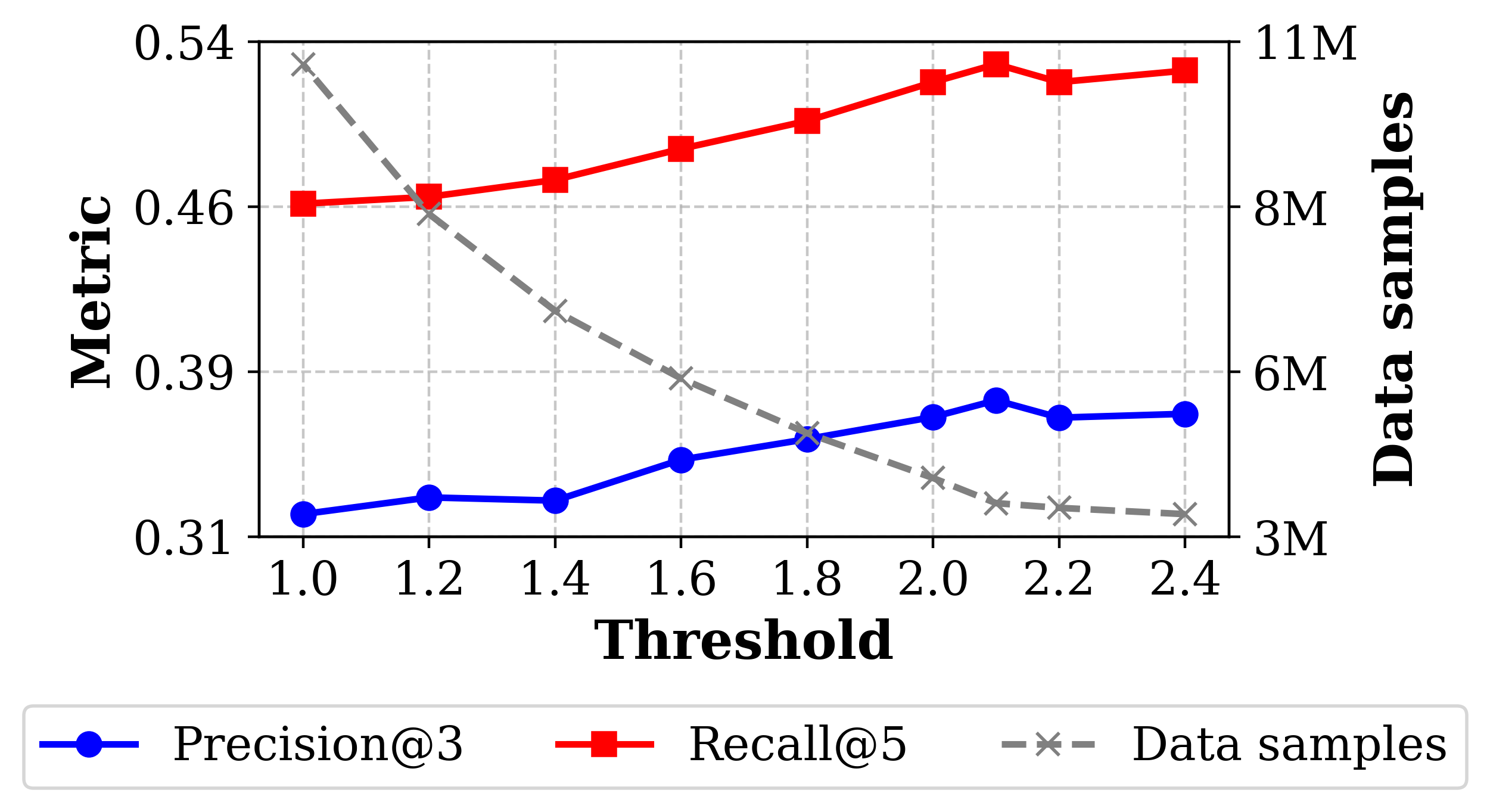}
\vspace{-1.5mm}
\caption{Performance over varying filtering thresholds. The numbers of data samples are shown in millions.}
\label{fig:exp_threshold_perf}
\vspace{-2.5mm}
\end{figure}

Figure~\ref{fig:exp_threshold_perf} illustrates the impact of relevance-based data filtering on model performance. We filter low-quality query-document pairs from raw click logs, varying the threshold from 1.0 to 2.2. As the threshold increases, the training dataset size decreases from 10.99M to 3.95M samples. Performance peaks at 2.1, which we adopt as the optimal setting. Beyond this point, aggressive filtering degrades results due to insufficient training signals. It confirms that data quality is critical for learning robust query-to-DocID mappings.

\section{Annotation Guidelines for Human Evaluation}
\label{app:human_eval}
Annotators performed a blind side-by-side comparison to assess the overall quality of top-ranked search result. We used a 5-point Likert scale designed to holistically capture both information relevance and source authority. The detailed criteria are as follows:

\vspace{0.5mm}
\begin{itemize}[leftmargin=*,topsep=0pt,itemsep=-1ex,partopsep=2ex,parsep=1ex]
\vspace{0.5mm}
\item \textbf{5 (Excellent)}: The document fully satisfies the user's information needs with comprehensive and accurate content. The source is highly authoritative (e.g., official government website, public institution, or major corporation). This is the ideal result for the query.
\item \textbf{4 (Good)}: The document provides high-quality and relevant information that effectively addresses the need. The source is trustworthy but lacks top-tier official status (e.g., well-maintained expert blog, reputable news organization, or major community platform). The content is trustworthy, but the source itself lacks official status.
\item \textbf{3 (Fair)}: The document partially satisfies the needs, but coverage is limited. The source has low or indeterminate authority (e.g., personal blog, forum post, or small commercial site).
\item \textbf{2 (Poor)}: The document is marginally relevant with very limited information, requiring additional searches to satisfy the need. The source quality is typically low.
\item \textbf{1 (Bad)}: The document is irrelevant and very low quality (e.g., spam, heavy ads), or inaccessible (e.g., broken links).
\end{itemize}

\noindent
In addition to the 1-5 rating, annotators provided a direct preference judgment ("Which result is better overall?") for each query to calculate the win ratio used in our main analysis.

\section{Role of Continued Pre-Training in Search Domain}
\label{app:cpt}
Continued pre-training (CPT) on domain-specific corpora is essential for adapting LLMs to search applications, bridging the gap between generic linguistic knowledge and relevance-oriented retrieval tasks. Without CPT, LLMs often remain biased toward generic next-token distributions, failing to capture implicit user relevance signals that imply retrieval and ranking. \citet{ye-etal-2025-best} demonstrate that applying CPT on a click-stream corpus, structured as \textit{(query, clicked title, summary)} triples, effectively aligns models with user interaction patterns and domain-specific semantics before fine-tuning. Following CPT, their framework fine-tunes the LLM with a pairwise ranking loss and subsequently distills knowledge into a BERT-based encoder via a hybrid loss combining absolute score regression and relative ranking margins. Their experiments confirm that CPT yields statistically significant gains, boosting NDCG@5 from 0.8709 to 0.8793. Importantly, this CPT-based pipeline was also deployed in a commercial search product, highlighting its practical necessity beyond controlled benchmarks. These findings establish CPT not as an optional enhancement, but as an indispensable phase for ensuring domain alignment and performance in real-world search systems.

\end{document}